\documentclass[12pt]{iopart}

\usepackage{graphicx}% Include figure files
\usepackage{dcolumn}% Align table columns on decimal point
\usepackage{bm}% bold math
\usepackage{epsfig}

\begin{document}

%%%%%%%%%%%%%%%      define  environment     %%%%%%%%%%%%%%%%%%%%%
\def\0#1#2{\frac{#1}{#2}}
\def\bct{\begin{center}} \def\ect{\end{center}}
\def\beq{\begin{equation}} \def\eeq{\end{equation}}
\def\bea{\begin{eqnarray}} \def\eea{\end{eqnarray}}
\def\nnu{\nonumber}
\def\n{\noindent} \def\pl{\partial}
\def\g{\gamma}  \def\l{\lambda} \def\e{\varepsilon} \def\o{\omega}
\def\s{\sigma}  \def\b{\beta} \def\p{\psi} \def\r{\rho}
%%%%%%%%%%%%%%%%%%%%%%%%%%%%%%%%%%%%%%%%%%%%%%%%%%%%%%%%%%%%%%%%

\title{Bose-Einstein condensation and chiral phase transition in linear sigma model}

\author{Song~Shu\footnote[1]{shus@iopp.ccnu.edu.cn} and Jia-Rong~Li}

\address{Institute of Particle Physics, Hua-Zhong Normal
University, Wuhan 430079, P. R. China}

\begin{abstract}
With the linear sigma model, we have studied Bose-Einstein
condensation and the chiral phase transition in the chiral limit
for an interacting pion system. A $\mu-T$ phase diagram including
these two phenomena is presented. It is found that the phase plane
has been divided into three areas: the Bose-Einstein condensation
area, the chiral symmetry broken phase area and the chiral
symmetry restored phase area. Bose-Einstein condensation can
happen either from the chiral symmetry broken phase or from the
restored phase. We show that the onset of the chiral phase
transition is restricted in the area where there is no
Bose-Einstein condensation.
\end{abstract}  \maketitle

\section{Introduction}
In relativistic heavy ion collisions, large amount of pions can be
created in small space-time regions. Theoretically, it is possible
for these pions to condensate into the zero momentum and form
Bose-Einstein condensation(BEC) at certain temperatures and
densities. For this problem, there are some discussions in the
literature, for example, the BEC of pions in the non-equilibrium
process~\cite{a,b}, the pion condensation in thermal equilibrium
incorporating boundary effects~\cite{c}, and the search for the
``cold spots" in multipion system~\cite{d,e}, etc. These studies
show a great interest on this issue. On the other hand, in high
energy physics, the chiral symmetry which is spontaneously broken
at low temperature is expected to be restored at high temperature
and/or density~\cite{f}. Long before, Kapusta and Haber had
discussed the relation between BEC and the spontaneous symmetry
breaking by using different scalar models~\cite{g,h}. Their main
interest, however, was focused on the physics of the electroweak
system. Here in the strong interacting pion system, one may wonder
whether there are some relations between the chiral symmetry
restoration and BEC.

Our motivation in this paper is to discuss the relation between
these two phenomena in the pion system and find the phase diagram
of them through a model study. As far as we know, this topic has
not been much discussed in the literature. In~\cite{i}, the chiral
symmetry restoration is discussed in the interaction pion gas
through virial expansion, where BEC is only discussed in the case
of free pion gas. Here we will discuss BEC and the chiral phase
transition in the interacting pion system with the linear sigma
model. The linear sigma model has been extensively used to discuss
the chiral symmetry restoration at finite
temperature~\cite{j,k,l}. In our discussion, to address BEC and
the chiral phase transition simultaneously, we will consider a
finite pion density as well as the temperature. The pion chemical
potential can be introduced in different ways. Theoretically, at
given conditions when the pion number can be regarded as being
fixed~\cite{m}, one could introduce the pion number density.
However, generally speaking, pion number is not a conserved
quantity and one should find a certain conserved charge of the
system. For the linear sigma model which has the isospin
invariance, one can introduce a chemical potential associated with
the isospin charge. For example, in~\cite{n}, the pion isospin
chemical potential is introduced by the third component of the
isospin charge and applied to the discussions of the thermal pion
mass within the chiral perturbation theory. In our discussion, the
isospin chemical potential of pions is introduced to the linear
sigma model in the same way.

The organization of the present paper is as follows. In section 2
the effective potential for the linear sigma model is calculated
within the Cornwall, Jackiw, Tomboulis resummation scheme at
finite temperature and finite pion density in the Hartree
approximation. The gap equations are derived by the stationary
condition, which form the basis of our numerical calculation. In
section 3, the BEC and the chiral phase transition are studied by
a numerical calculation. We discuss their $\mu-T$ phase diagrams
and give the physical analysis. The last section comprises a
summary of our results.

\section{CJT formalism and effective Lagrangian}
In studies of phase transitions the effective potential is an
important and popular theoretical tool. It is defined through an
effective action which is the generating functional of the one
particle irreducible graphs. A generalized version is the
effective potential for composite operators introduced by
Cornwall, Jackiw and Tomboulis(CJT)~\cite{o}. This formalism
assumes that the effective potential $V(\phi, G)$ depends not only
on $\phi(x)$, a possible expectation value of the quantum field
$\Phi(x)$, but also on $G(x,y)$, a possible expectation value of
the time-ordered product $T\Phi(x)\Phi(y)$. Physical solutions
demand minimization of the effective potential with respect to
both $\phi$ and $G$, which means \bea \0{dV(\phi,G)}{d\phi}=0, \ \
\ \ \ \ \0{dV(\phi,G)}{dG}=0. \eea This formalism was originally
written at zero temperature. Then it has been extended to finite
temperature occasion by Amelino-Camelia and Pi for investigations
of the effective potential of the $\l\phi^4$ theory~\cite{p}. It
has been also used in the linear sigma model for discussing the
chiral phase transition at finite temperature~\cite{j}. At finite
density as well as finite temperature, we will employ this
formalism to discuss both BEC and the chiral phase transition with
the linear sigma model.

For the linear sigma model, here we will ignore the fermion
sector. The Lagrangian can be written as \bea {\cal
L}=\012(\pl\bf\s)^2&+&\012(\pl\vec\pi)^2-\012m^2\s^2-\012m^2\vec\pi^2
\nnu
\\ &-&\0\l{24}(\s^2+{\vec\pi}^2)^2-\e\s , \eea
where $\s$ and ${\vec\pi}$ are the sigma field and the three pion
fields ($\pi_1, \pi_2, \pi_3$) respectively. The explicit chiral
symmetry breaking term is $\e=f_\pi m^2_\pi$, where $f_\pi=93MeV$
is the pion decay constant. The coupling constant $\l$ and
negative mass parameter $m^2$ of the model are chosen to be
$\l=3(m^2_\s-m^2_\pi)/f^2_\pi$ and $-m^2=(m^2_\s-3m^2_\pi)/2>0$.
In our discussion, we only consider the case in the chiral limit
$\e=0$, then at zero temperature the pion mass is $m_\pi=0$ and
the sigma mass is taken as $m_\s\approx 600MeV$.

The conventional treatment of spontaneous chiral symmetry breaking
of the linear sigma model is to shift the sigma field as
$\s\to\s+\phi$, where $\phi$ is expectation value of the sigma
field and the order parameter of the chiral phase transition. Then
the Lagrangian results as \bea {\cal
L}=\012(\pl\vec\pi)^2&+&\012(\pl\s)^2-\012(m^2+\0\l2\phi^2)\s^2
-\012(m^2+\0\l6\phi^2)\vec\pi^2 \nnu \\ &-&\012m^2\phi^2
-\0\l{24}\phi^4
-\0\l{24}(\s^2+\vec\pi^2)^2-\0\l6\phi\s^3-\0\l6\phi\s\pi^2. \eea
In the following discussion, we will adopt Hartree approximation
which means only the bubble diagrams with four-point vertex need
to be calculated. So the last two terms in the Lagrangian are
neglected~\cite{j}.

The chemical potential is introduced through the third component
of isospin charge. The Lagrangian (3) is invariant under the
isospin transformation, therefore according to Noether's theorem
the conserved current is $\vec V_\mu =\vec\pi\times\pl_\mu\vec\pi$
and the conserved isospin charge is $\vec V_0
=\vec\pi\times\pl_0\vec\pi.$ Then the partition function of the
system can be written as \bea Z=\int
[d\Pi_{\vec\pi}][d\Pi_\s][d\vec\pi][d\s]exp\left[i\int d^4x({\cal
L}+\vec\mu_I\cdot\vec V_0)\right], \eea where $\Pi_{\vec\pi}$ and
$\Pi_\s$ are the conjugate momenta of $\vec\pi$ and $\s$. The
isospin chemical potential $\vec\mu_I=(0, 0, \mu)$, in which $\mu$
is the third component of isospin chemical potential, is
introduced associated with $\vec V_0$~\cite{n}, thus \bea
\vec\mu_I\cdot\vec V_0=\mu (\pi_1\pl_0\pi_2-\pi_2\pl_0\pi_1). \eea
Here it is more convenient to work with fields possessing
well-defined charges: \bea \pi_-\equiv\01{\sqrt2}(\pi_1+i\pi_2), \
\ \pi_+\equiv\01{\sqrt2}(\pi_1-i\pi_2),\ \ \pi_0\equiv\pi_3. \eea
Integrating out the conjugate momenta in equation (4) and using
the redefined pion fields, one obtains \bea
Z=\int[d\pi_+][d\pi_-][d\pi_0][d\s]exp\left[i\int d^4x{\cal
L}_{eff}\right], \eea where ${\cal L}_{eff}$ is the effective
Lagrangian of the system and can be written as, \bea &{\cal
L}&_{eff}=|\pl_I\pi|^2-(m^2+\0\l6\phi^2)|\pi|^2+\012(\pl\pi_0)^2+\012(\pl\s)^2
 \nnu \\
&-&\012(m^2+\0\l6\phi^2)\pi^2_0-\012(m^2+\0\l2\phi^2)\s^2-\012m^2\phi^2-\0\l{24}\phi^4
\nnu \\
&-&\0\l{24}\lbrack\s^4+4|\pi|^4+\pi_0^4+4\s^2|\pi|^2+2\s^2\pi_0^2+4|\pi|^2\pi_0^2\rbrack
\eea with $|\pl_I\pi|^2=(\pl_\mu+i\mu\delta_{\mu 0}
)\pi_+(\pl^\mu-i\mu\delta^{\mu 0})\pi_-$ and $|\pi|^2=\pi_+\pi_-$.
It should be noted that the chemical potential $\mu$ has entered
the lagrangian in $|\pl_I\pi|^2$. The following calculations and
discussions will be based on this effective Lagrangian.

\section{Gap equations and thermodynamic potential}
Within the imaginary time formalism of finite temperature field
theory, we have \bea
\int\0{d^4k}{(2\pi)^4}f(k)\to\01{\b}\sum_n\int\0{d^3\bf
k}{(2\pi)^3}f(i\o_n,{\bf k})  \equiv\int_\b f(i\o_n, {\bf k}),
\eea where $\b$ is the inverse temperature, $\b=1/T$; the
integration over the time component $k_0$ has been replaced by a
summation over discrete frequencies. For boson there are
$\o_n=2\pi nT$ and $n=0, \pm1, \pm2, \cdots$.  For the sake of
simplicity in what follows, a shorthand notation is used to denote
the integration and the summation.

According to the Lagrangian (8), the tree level inverse
propagators of $\s$, $\pi_0$ and $\pi_\pm$ can be written
respectively as
\bea D^{-1}_\s&=&\o^2_n+{\bf k}^2+m^2+\0\l2\phi^2,  \\
D^{-1}_0&=&\o^2_n+{\bf k}^2+m^2+\0\l6\phi^2, \\
D^{-1}&=&(\o_n+i\mu)^2+{\bf k}^2+m^2+\0\l6\phi^2. \eea The
equation (12) represents the inverse propagator of $\pi_+$ and
$\pi_-$. As the summation in equation (9) is symmetric over $n$
from $-\infty$ to $+\infty$, $\o_n+i\mu$ and $\o_n-i\mu$ are
equivalent in describing the propagators of $\pi_+$ and $\pi_-$.
Here we just choose $\o_n+i\mu$ to describe both $\pi_+$ and
$\pi_-$, and write the two propagators in the same form as in
equation (12). According to \cite{j}, we can write down the
corresponding effective potential at finite temperature as \bea
V(&\phi&, G)=\012m^2\phi^2+\0\l{24}\phi^4+\012\int_\b\ln G^{-1}_\s
+\012\int_\b \left[D^{-1}_\s G_\s-1\right]+\012\int_\b\ln G^{-1}_0
\nnu \\ &+&\012\int_\b\left[D^{-1}_0G_0-1\right] +\int_\b\ln
G^{-1} +\int_\b\left[D^{-1}G-1\right]+V_2(\phi, G), \eea where
$G_\s, G_0$ and $G$ are the full propagators of $\s, \pi_0$ and
$\pi_\pm$ respectively. They are determined by the stationary
condition (1). The chemical potential is included in the full
propagator. $V_2(\phi,G)$ represents the infinite sum of the two
particle irreducible vacuum graphs. However, in CJT formalism at
the Hartree approximation we need only to calculate the ``double
bubble" diagrams with four-point vertex and treat each loop line
as the full propagator~\cite{j,o}. Therefore, $V_2$ can be written
as \bea V_2(\phi, G)&=&\0\l8\left[\int_\b
G_\s\right]^2+\0\l3\left[\int_\b
G\right]^2+\0\l8\left[\int_\b G_0\right]^2 \nnu \\
&+&\0\l6\int_\b G_\s\int_\b G+\0\l{12}\int_\b G_\s\int_\b G_0
+\0\l6\int_\b G\int_\b G_0. \eea

By minimizing the effective potential $V(\phi, G)$ with respect to
the full propagators, we get the following set of nonlinear gap
equations, \bea G^{-1}_\s&=&D^{-1}_\s+\0\l2\int_\b
G_\s+\0\l3\int_\b G+\0\l6\int_\b G_0, \\
G^{-1}_0&=&D^{-1}_0+\0\l2\int_\b G_0+\0\l6\int_\b
G_\s+\0\l3\int_\b G,  \\ G^{-1}&=&D^{-1}+\0{2\l}3\int_\b
G+\0\l6\int_\b G_\s+\0\l6\int_\b G_0 . \eea Furthermore we take
the similar ansatz for the full propagators as in \cite{j},
\bea G^{-1}_\s&=&\o^2_n+{\bf k}^2+M^2_\s, \\
G^{-1}_0&=&\o^2_n+{\bf k}^2+M^2_0, \\ G^{-1}&=&(\o_n+i\mu)^2+{\bf
k}^2+M^2, \eea where the effective masses $M_\s, M_0$ and $M$ have
been introduced for $\s, \pi_0$ and $\pi_\pm$ respectively.
Substituting them into the above gap equations, we obtain a set of
effective mass gap equations, \bea
M^2_\s&=&m^2+\0\l2\phi^2+\0\l2\int_\b
G_\s+\0\l3\int_\b G+\0\l6\int_\b G_0, \\
M^2_0&=&m^2+\0\l6\phi^2+\0\l2\int_\b G_0+\0\l6\int_\b
G_\s+\0\l3\int_\b G, \\ M^2&=&m^2+\0\l6\phi^2+\0{2\l}3\int_\b
G+\0\l6\int_\b G_\s+\0\l6\int_\b G_0. \eea Accordingly the
effective potential can be written as \bea V(\phi,
M)&=&\012m^2\phi^2+\0\l{24}\phi^4+\012\int_\b\ln
G^{-1}_\s \nnu \\
&-&\012\int_\b(M^2_\s-m^2-\0\l2\phi^2)G_\s+\012\int_\b\ln G^{-1}_0
\nnu \\ &-&\012\int_\b(M^2_0-m^2-\0\l6\phi^2)G_0+\int_\b\ln G^{-1}
\nnu \\ &-&\int_\b(M^2-m^2-\0\l6\phi^2)G+V_2(\phi, M). \eea By
minimizing the potential with respect to the order parameter
$\phi$, we obtain one more equation, \bea
m^2\phi+\0\l6\phi^3+\0{\l\phi}2\int_\b G_\s+\0{\l\phi}6\int_\b
G_0+\0{\l\phi}3\int_\b G=0. \eea From equations (21)---(23) and
(25), the effective masses and order parameter can be solved
self-consistently at given temperature and chemical potential.
There are some discussions concerning the renormalization on the
CJT formalism in different models~\cite{p,q,r}. Recent
investigation about the renormalization of the O(N) linear sigma
model has been addressed in \cite{k}. In our discussion, the
divergent parts of the integrals are neglected as is done in
\cite{j}. And we take the finite parts of the integrals as
follows:  \bea &&\int_\b\ln G^{-1}=\int_\b\ln[(\o_n+i\mu)^2+{\bf
k}^2+M^2] \nnu
\\
&=&\01{\b}\int\0{d^3{\bf
k}}{(2\pi)^3}\left[\ln(1-e^{-\b(\o+\mu)})+\ln(1-e^{-\b(\o-\mu)})\right],
\\ &&\int_\b G=\int_\b\01{(\o_n+i\mu)^2+{\bf k}^2+M^2} \nnu \\ &=&\int\0{d^3{\bf
k}}{(2\pi)^3}\01{2\o}\left[\01{e^{\b(\o+\mu)}-1}+\01{e^{\b(\o-\mu)}-1}\right],
\eea where $\o=\sqrt{{\bf k}^2+M^2}$. The integration over $G_0$
and $G_\s$ will be dealt with in the same way. With the chiral
limit ($\e=0$), the corresponding coupling constant and the
negative mass parameter can be taken as $\l\approx 125$ and
$-m^2\approx 1.8\times 10^5MeV^2$.

In the discussion of the thermodynamic system, the effective
potential $V$ is equivalent to the thermodynamic potential
$\Omega$. Thus we get \bea
\Omega&=&\012m^2\phi^2+\0\l{24}\phi^4+T\int\0{d^3{\bf
k}}{(2\pi)^3}\ln(1-e^{-\b\o_\s}) \nnu \\
&-&\012(M^2_\s-m^2-\0\l2\phi^2)\int\0{d^3{\bf
k}}{(2\pi)^3}\01{\o_\s}\01{e^{\b\o_\s}-1} \nnu \\
&+&T\int\0{d^3{\bf k}}{(2\pi)^3}\ln(1-e^{-\b\o_0})
-\012(M^2_0-m^2-\0\l6\phi^2) \int\0{d^3{\bf
k}}{(2\pi)^3}\01{\o_0}\01{e^{\b\o_0}-1} \nnu \\ &+&T\int\0{d^3{\bf
k}}{(2\pi)^3}\left[\ln(1-e^{-\b(\o+\mu)})+\ln(1-e^{-\b(\o-\mu)})\right]
\nnu \\ &-&(M^2-m^2-\0\l6\phi^2)\int\0{d^3{\bf
k}}{(2\pi)^3}\01{2\o}
\left[\01{e^{\b(\o+\mu)}-1}+\01{e^{\b(\o-\mu)}-1}\right]+\Omega_2,
\eea where $\Omega_2=V_2$, $\o_\s=\sqrt{{\bf k}^2+M^2_\s},
\o_0=\sqrt{{\bf k}^2+M^2_0}$ and $\o=\sqrt{{\bf k}^2+M^2}$.
According to the relation \bea \rho=-\0{\pl\Omega}{\pl\mu}, \eea
and equation (23), we can get the net charge density \bea
\r=\int\0{d^3{\bf
k}}{(2\pi)^3}\left[\01{e^{\b(\o-\mu)}-1}-\01{e^{\b(\o+\mu)}-1}\right].
\eea This expression of density seems very similar to that of
ideal gas, but actually they are different. Because here the
effective mass $M$ in $\o$ is a function of temperature and will
be determined self-consistently by the gap equations.

\section{BEC and chiral phase transition}
Now we are in a position to discuss BEC and the chiral phase
transition. We will study BEC first. It is known that when
$\mu=M$, BEC occurs. The equation (30) should be written as
\bea \r=\r_0+\r^*(\b,\mu=M), \nnu \\
\r^*(\b,\mu=M)=\int\0{d^3{\bf
k}}{(2\pi)^3}\left[\01{e^{\b(\o-M)}-1}-\01{e^{\b(\o+M)}-1}\right],
\eea where $\r_0$ represents the charge density of the
zero-momentum state~\cite{p1,s}. Let us determine the point at
which $\mu=M$ is reached. From equation (30), when $\r$ is fixed,
by solving the gap equations (21)---(23) and equation (25), we
find both $\mu$ and $M$ are functions of $T$. It can be plotted in
figure 1. We can see that with $T$ decreasing, $M$ decreases first
and then increases, while $\mu$ keeps increasing quickly and
approaches $M$. At certain temperature, $\mu$ catches up with $M$
and BEC happens. From equation (31), we know the critical
temperature $T_c$ is determined implicitly by the equation \bea
\r=\r^*(\b_c,\mu=M). \eea When $T\leq T_c$, the system goes into
BEC phase. The equation (31) will be solved together with the gap
equations with $\r$ fixed. We find $\mu$ and $M$ are still
functions of $T$ and $\mu(T)=M(T)$. They both decrease with
temperature decreasing, which is indicated in figure 1.
\begin{figure}[tbh]
\begin{center}
\includegraphics[width=220pt,height=130pt]{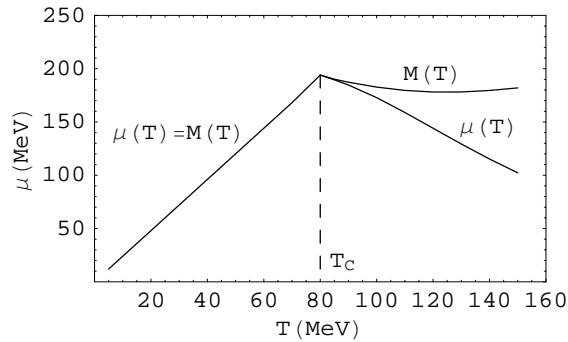}
\end{center}
\caption{$\mu$ and $M$ as functions of $T$ at the fixed total
charge density $\r=0.06fm^{-3}$. BEC happens at $T_c=80MeV$.}
\end{figure}
\begin{figure}[th]
\begin{center}
\includegraphics[width=220pt,height=130pt]{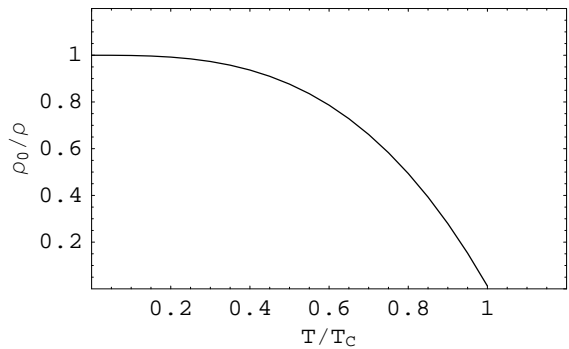}
\end{center}
\caption{The ratio $\r_0/\r$ as a function of the ratio $T/T_c$ at
$\r=0.06fm^{-3}$ and $T_c=80MeV$.}
\end{figure}

From equation (31) which is solved together with the gap equations
at fixed $\r$, we can get $\r_0$ as a function of $T$. This is
shown in figure 2. It is clear that when $\r$ is fixed, at
$T>T_c$, $\r_0=0$; at $T\leq T_c$, $\r_0$ starts to increase
quickly with temperature decreasing, which means large amount of
the charged pions will reside to the zero momentum state and form
BEC.

For different fixed total density $\r$, there will be different
values of critical temperature $T_c$ and chemical potential
$\mu(T_c)$, so one can study the $\mu-T$ phase diagram of BEC. In
solving the self-consistent equations, we observe that $T_c$ and
$\mu(T_c)$ depend on the order parameter $\phi$. In low
temperature regime, the order parameter $\phi$ is nonzero and
determined by equation (25). When it is solved together with the
gap equations (21)---(23) and equation (32), the critical
temperature of BEC at a given density can be determined. The
$\mu-T$ phase diagram for BEC can be given accordingly as shown in
figure 3. The dividing line which separates BEC and the normal
phase at $\phi\neq 0$ is curve AC. At $T>T_C$ (the temperature at
the point C), there is no solution with $\phi\neq 0$ for the onset
of BEC. Because in the high temperature regime the chiral symmetry
is restored and $\phi=0$. At $\phi=0$ the phases dividing line is
curve DF as shown in figure 3. We can see that BEC happens in
chiral symmetry broken state and restored state at low temperature
and high temperature respectively.

In the regime of $T_B<T<T_C$, it seems that BEC can happen at both
$\phi\neq 0$ and $\phi=0$. However, if we calculate thermodynamic
potential from equation (28), we find under the same temperature
the thermodynamic potential of $\phi\neq 0$ has lower values than
that of $\phi=0$, which means at $T_B<T<T_C$, BEC would happen
along the curve BC. Then the final $\mu-T$ phase diagram of BEC is
determined and shown in figure 4. In low temperature regime BEC
phase and the normal phase are divided by curve AC, while in high
temperature regime, the phases are divided by curve EF. The whole
phase plane has been divided into two phase areas. The area above
the curve belongs to the BEC phase, while the area below the curve
represents the normal phase.
\begin{figure}[tbh]
\begin{center}
\includegraphics[width=220pt,height=130pt]{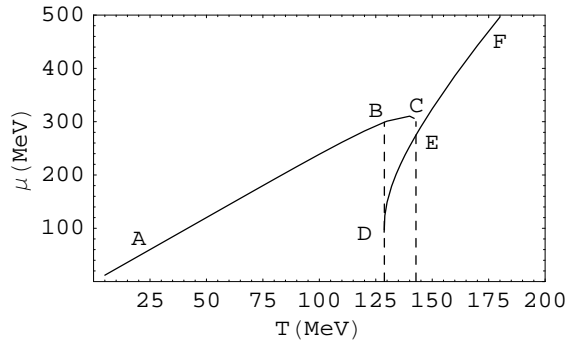}
\end{center}
\caption{The phase diagram of $\mu$ versus $T$ for BEC. (AC and DF
are the dividing lines which separate BEC and normal phases in low
temperature regime and high temperature regime respectively.)}
\end{figure}
\begin{figure}[tbh]
\begin{center}
\includegraphics[width=220pt,height=130pt]{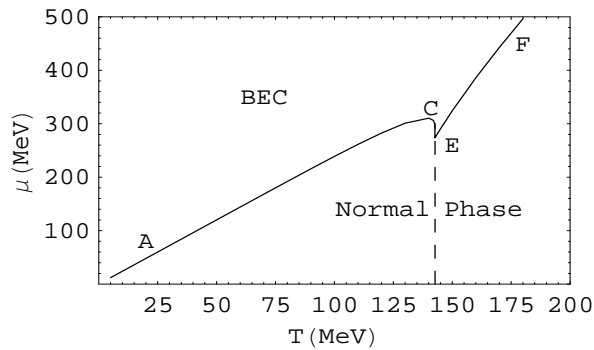}
\end{center}
\caption{The $\mu-T$ phase diagram for BEC with no ambiguity.}
\end{figure}

Now we will discuss the chiral phase transition in this system. By
using the linear sigma model, there are many discussions on the
chiral phase transition at finite temperature~\cite{j,k,l}. As we
have considered finite pion density as well as finite temperature,
the calculation becomes more involved. However, the chiral phase
transition is still characterized by the order parameter $\phi$.
In our discussion, $\phi$ is not only temperature dependent but
also chemical potential dependent. In \cite{j}, the chiral phase
transition has been discussed at finite temperature with the
linear sigma model. By using CJT resummation scheme in Hartree
approximation, the transition is found to be first order in the
chiral limit. If we take $\mu=0$, we can reproduce their result.
In the case of $\mu\neq 0$, we find the chiral phase transition is
still first order. This can be seen if we take a look at $\Omega$
as a function of $\phi$. As the gap equations (21)---(23) can be
solved at certain $T$ and $\mu$, from equation (28), the
thermodynamical potential $\Omega$ can be determined. The $\phi$
dependence of $\Omega$ can be plotted as shown in figure 5. In the
transition, it can be seen that there are two minima of $\Omega$
which correspond to the stable and meta-stable states at certain
$T$ and $\mu$. If $T$ is fixed, by increasing $\mu$, the lower
minimum of $\phi\neq 0$, which stays in a stable state, will be
lifted up. When it becomes higher than the $\phi=0$ minimum, it
changes to a meta-stable state from the stable state. In other
words, the $\phi=0$ minimum becomes a stable state which means
chiral symmetry is restored. It is clear that the chiral phase
transition is first order. The transition is usually defined at
the point where the two minima become degenerate. In figure 5,
when $T=160MeV$ and $T=180MeV$, we can see that the transitions
happen at $\mu=237MeV$ and $\mu=100MeV$ respectively. According to
this way, the $\mu-T$ phase diagram for the chiral phase
transition can be plotted as shown in figure 6.
\begin{figure}[t]
\begin{center}
\includegraphics[width=220pt,height=130pt]{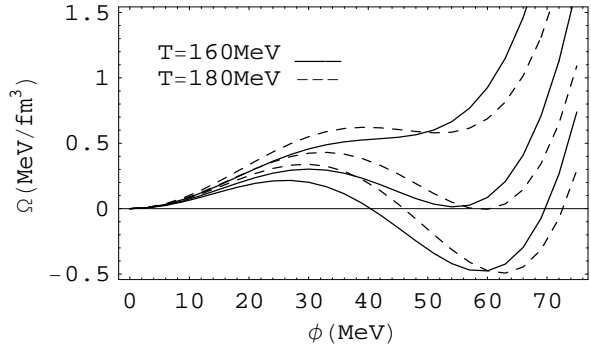}
\end{center}
\caption{$\Omega$ as functions of $\phi$. Solid lines stand for
$T=160MeV$, and the chemical potentials for different lines are
$260MeV, 237MeV$ and $220MeV$ respectively (from top to bottom).
Dashed lines stand for $T=180MeV$, and the chemical potentials are
$140MeV, 100MeV$ and $60MeV$ respectively (also from top to
bottom).}
\end{figure}
\begin{figure}[h!]
\begin{center}
\includegraphics[width=220pt,height=130pt]{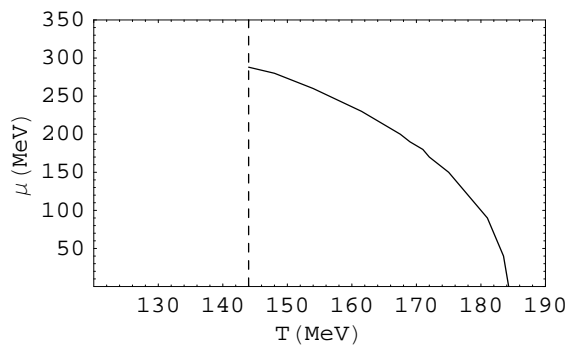}
\end{center}
\caption{The phase diagram of $\mu$ versus $T$ for the chiral
phase transition. When $T>144MeV$, the area above the curve is the
restored phase; the area below the curve is the broken phase.}
\end{figure}

From figure 6, we can see that the transition temperature is
$184MeV$ for $\mu=0$, which was obtained in \cite{j}. Here when
$\mu\ne 0$, the transition temperature will decrease with the
chemical potential increasing. At certain temperature ($T\approx
144MeV$), the phase curve of the chiral phase transition will
terminate, because at this time $\mu=M$ and BEC will happen. As a
result the chiral phase transition can not take place further at
lower temperatures. When $T>144MeV$, the area above the curve is
the chiral symmetry restored phase. However, the phase diagram is
not complete, because we do not know what the phase is when
$T<144MeV$. Now let us recall the $\mu-T$ phase diagram of BEC
(figure 4). If we combine the two phase diagrams (figure 4 and
figure 6) into one figure, we can see the different phase areas
clearly. This is shown in figure 7. When the chiral phase
transition terminates, it means that the system will go into BEC
phase. Furthermore, the normal phase area in figure 4 is divided
into the chiral symmetry broken phase and the chiral symmetry
restored phase. When the system is in a chiral symmetry broken
phase, if we decrease temperature with density fixed, BEC will
occur at ceratin temperature; if we increase temperature, the
system may go into a chiral symmetry restored phase. In the
restored phase, at relatively low densities, if we decrease
temperature, the chiral phase transition may take place; at
relatively high densities, if we decrease temperature with density
fixed, BEC will take place.
\begin{figure}[!h]
\begin{center}
\includegraphics[width=220pt,height=130pt]{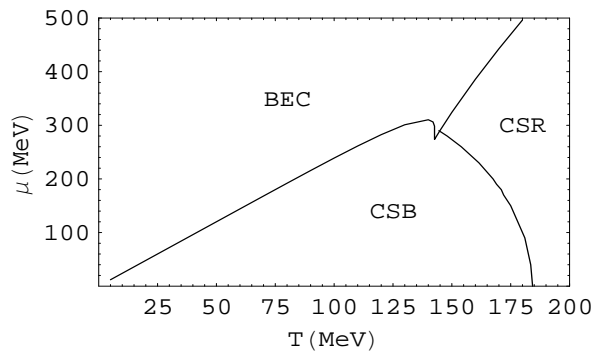}
\end{center}
\caption{The combined phase diagram of $\mu$ versus $T$ for BEC
and the chiral phase transition. The CSB and CSR stand for chiral
symmetry broken phase and chiral symmetry restored phase
respectively.}
\end{figure}

We noted that in \cite{h}, BEC and the spontaneous symmetry
breaking in interacting Bose system have been discussed with the
O(N) model. In their discussion, a chemical potential $\mu$
associated with O(2) symmetry is introduced when N is even. There
are $N/2$ such chemical potentials for $N$ scalars. For only one
chemical potential and the large-N limit, their result shows that
the transition to broken symmetry phase is just the onset of BEC
in the the ultra-relativistic regime. This is because the scalar
field associated with $\mu$ acquires a nonzero vacuum expectation
value at the point that $\mu=M$ is reached when temperature is
lowered. Thus the symmetry, which is actually the O(2) symmetry,
is broken at this time. In our discussion, we have studied the
chiral symmetry restoration and BEC in the O(4) linear sigma
model. In this model, the four scalar fields have real physical
meaning, which are three pion fields and one sigma field. The
chiral symmetry breaking in the vacuum is chosen along the $\s$
direction. The chiral phase transition is characterized by the
order parameter $\phi$. The chemical potential here is associated
with the pion fields. The BEC means the BEC of the charged pions.
The relation between BEC and the chiral phase transition can be
illustrated by the $\mu-T$ phase diagram as indicated in figure 7.

\section{Conclusion}
In summary, we have discussed Bose-Einstein condensation and the
chiral phase transition for the linear sigma model in the chiral
limit. The effective potential of the linear sigma model is
calculated within the CJT resummation scheme in the Hartree
approximation at finite temperature and finite pion density. By
solving the self-consistent equations we have discussed BEC in the
interacting pion system. The $\mu-T$ phase diagram of BEC is
presented. At relatively low temperature, BEC happens in chiral
symmetry broken state, while at relatively high temperature, BEC
happens in chiral symmetry restored state. The phase plane has
been divided into the BEC phase and the normal phase. Then we have
discussed the chiral phase transition and also plotted the $\mu-T$
phase digram of the chiral phase transition. By comparing it with
the BEC phase diagram, we find the chiral phase transition happens
in the normal phase and further divided the normal phase into the
chiral symmetry broken phase and restored phase. So the whole
phase plane has been divided into three phase areas: the BEC phase
area, the chiral symmetry broken phase area and the chiral
symmetry restored phase area. BEC can happen either from the
chiral symmetry broken phase or from the chiral symmetry restored
phase. When the system goes into BEC phase, there is no chance for
the onset of the chiral phase transition.

\ack We would like to thank Professor Ji-Sheng Chen and Professor
De-Fu Hou for their help. This work was supported in part by the
National Natural Science Foundation of China with No. 90303007.

\end{document}